\documentclass[epj]{svjour}
\usepackage{graphics,color}
\usepackage[numbers]{natbib}
\usepackage{amsmath}
\usepackage{amssymb}
\usepackage{ulem}
\usepackage{multirow}
\usepackage{subcaption}
\usepackage{cite}
\usepackage{fontenc}
\usepackage{bm}
\usepackage{dcolumn}
\usepackage{times}
\usepackage{pdflscape}
\usepackage{epsfig}
\usepackage{rotating}
\usepackage{tabularx}
\usepackage{natbib}
\usepackage{hyperref}
\usepackage{float}
\usepackage{soul,xcolor}
\setstcolor{red}   
\usepackage{cancel}
\usepackage{gensymb}
\usepackage{footnote}

\begin{document}
\title{Low-lying dipole strengths for probable $p$-wave one-neutron halos in the medium mass region}
%
\author{Manju\inst{1} \and Jagjit Singh\inst{2} \and Shubhchintak\inst{3} \and R. Chatterjee\inst{1}%
}                     
\offprints{Manju}
\mail {manju@ph.iitr.ac.in}
\institute{Department of Physics, Indian Institute of Technology Roorkee, Uttarakhand- 247667, India 
\and Nuclear Reaction Data Centre, Faculty of Science, Hokkaido University, Sapporo 060-0810, Japan
\and Physique Nucl\'{e}aire Th\'{e}orique et Physique Math\'{e}matique, C. P. 229, Universit\'{e} Libre de Bruxelles (ULB), B 1050 Brussels, Belgium}
\date{Received: date / Revised version: date}
%
\abstract{
The one-neutron halos lying in the island of inversion around $N=20$ has provided the podium, 
to study the variation of total low-lying dipole strength with the neutron separation energy. 
We study three  probable $p$-wave one-neutron halo candidates $^{31}$Ne and $^{34}$Na and $^{37}$Mg lying in the island of inversion.
A simple analytic model has been used for the calculation of the total low-lying dipole strength for the medium mass $p$-wave one-neutron halos. A correction factor to this analytical model has been estimated with a realistic Woods-Saxon potential.
A comparison of these analytic calculations has been made with the those performed by a finite-range distorted-wave Born approximation theory of the Coulomb dissociation. We also make an estimate of the one-neutron separation energies of $^{31}$Ne, $^{34}$Na and $^{37}$Mg.}
\maketitle
\section{Introduction}\label{intro}
The advent of new dedicated radioactive-ion beam (RIB) facilities around various parts of the globe has provided access to 
proton and neutron-rich exotic nuclei in and beyond the $pf$- and $sd$- shell regions. 
One of the most notable feature among these short lived exotic nuclei is the formation of proton and neutron halos \citep{Tanihata1985}, found near the proton 
and neutron drip lines, respectively. In the neutron-rich region, the neutron halos are not only found in the 
lighter mass nuclei but also possibly in the nuclei as heavy as Mg \citep{Kobayashi2014}. 
These neutron halos are characterized by the large spatial extension \citep{Tanihata1985}, large matter radius \citep{Hansen1995}, 
small neutron separation energies \citep{Tanihata1985,Hansen1995}, and concentration of electric dipole strength at low 
excitation energy \citep{Aumann1999, Nakamura2006, Catara1996}, in comparison to the neighboring isotopes. Nowadays, 
the reaction studies involving these relatively heavy neutron halos has gained considerable attention. The experiments with 
these medium mass halos reported the strong dipole strength at the continuum threshold \citep{Bracco1999} 
and this strength can  be linked to the favorable matching of the weakly bound state and the low-lying continuum state in the transition 
matrix element \citep{Bertulani1991, Bertulani1992, Nagarajan2005, Typel2005}. 

In the medium mass region, for the neutron-rich isotopes of Ne, Na and Mg with $N\approx20$ lying in island of inversion, 
the vanishing of shell gaps causes the mixing of normal and intruder neutron configurations which further has strong 
impact on the ground state properties of these nuclei ~\citep{Warburton1990, Orr1991}. 
For the present study we choose three different possible $p$-wave one-neutron halos $^{31}$Ne and $^{34}$Na and $^{37}$Mg. 
Recently, these possible $p$-wave one-neutron halos has gained attraction from theoretical as well as experimental points of view.

In the conventional shell model picture, the valence neutron in the ground state of $^{31}$Ne ($N=21$) is expected to present in $0f_{7/2}$ orbit. 
In the late nineties, the dominant $0f_{7/2}$ valence-neutron configuration expected from conventional shell ordering was ruled out in various theoretical calculations and was predicted to be in $1p_{3/2}$ \citep{Poves1994, Descouvemont1999}. This $p$-wave dominance was confirmed later at RIKEN by 
observing large Coulomb breakup cross section for $^{31}$Ne, which is indicative of a soft $E1$ excitation \citep{Nakamura2009}. 
This measurement triggered off several theoretical investigations which leads to the conclusion that the spin-parity of the ground
state of $^{31}$Ne is ${3/2}^-$ \citep{Hamamoto2010,Horiuchi2010,Takechi2012,Shubhchintak2014,Juhee2017} rather than ${7/2}^-$ as suggested by shell model. 
The measured values for the one-neutron separation energy ($S_n$) of $^{31}$Ne are very small, i.e., $0.29\pm1.64$\,MeV \citep{Jurado2007}, $0.06\pm0.41$\,MeV \citep{Gaudefroy2012} and $0.15^{+0.16}_{-0.10}$\,MeV 
via $1n$-removal reactions \citep{Nakamura2014}, where as the evaluated $S_n$ is $0.30\pm1.6$\,MeV \citep{Ouellet2013}. 

There is scarce experimental and theoretical information available over the ground state of possible one-neutron halo 
$^{34}$Na ($N=23$) and $^{37}$Mg ($N=25$) in comparison to $^{31}$Ne. 
Recently, similar to the patterns detected in $^{31}$Ne, evidences of $p$-wave halos in the $^{34}$Na \citep{Gaudefroy2012, Fortune2013,  Doornenbal2014, Singh2016} 
and $^{37}$Mg \citep{Kobayashi2014, Shubhchintak2015} were reported. 
The measured one-neutron separation energy of $^{34}$Na is $0.17\pm50$\,MeV \citep{Gaudefroy2012} and its
evaluated value is $0.80\pm0.008$\,MeV \citep{Nica2012} whereas in Ref.~\citep{Fortune2013}, an upper limit of $0.2$\,MeV has been suggested for $S_n$. For $^{37}$Mg, 
the mass systematics suggests its $S_n$ to be $0.16\pm0.68$\,MeV \citep{Wang2012}. 
Coulomb dissociation measurements suggests it to be $0.22^{+0.12}_{-0.09}$\,MeV \citep{Kobayashi2014} and 
the value based on this measurement extracted with a finite-range distorted-wave Born approximation (FRDWBA) theory to be $0.35\pm0.06$\,MeV \citep{Shubhchintak2015}. 

In the context of large ambiguities present over the $S_n$values for these three $p$-wave one-neutron halos, 
we present the scaling of $S_n$ with the peak of dipole strengths by using an analytic model \citep{Nagarajan2005}.
Similar work have also been reported in Refs. \citep{Bertulani1991, Bertulani1992, Typel2005}.
In fact, in this simple analytic model the estimate of the total low-lying dipole strength for one-neutron halos in terms of their separation energy has been made and is 
implemented on lighter one neutron halo $^{19}$C \citep{Nagarajan2005}. The total strength obtained in this study was inversely proportional $S_n$. 

The main purpose of the present study is to investigate the analytic estimate of dipole strength distribution 
for the medium-mass nuclei $^{31}$Ne, $^{34}$Na and $^{37}$Mg lying in the island of inversion and to investigate the contribution of different transitions in the 
total strength. To compare the analytic results, we use the post form finite-range Distorted-wave Born approximation (FRDWBA) theory 
of the Coulomb dissociation \citep{Chatterjee2000}. Coulomb dissociation method has been widely used as a probe to study the multipole responses of the weakly-bound system. 
The dissociation cross-section can be linked to the electromagnetic strength \textit{B(E$\lambda$)}, 
which further provides information about the projectile ground state \citep{Bertulani1988, Hussein1991, Bertsch1991, Esbensen1992, Nakamura1994}. 
The peak positions of the dipole strength distribution could be used as a tool to predict the one-neutron separation energy of the concerned projectile.

The paper is organized as follows.
Section~\ref{MF} briefly describes the mathematical set up of the analytic model and the FRDWBA theory.
Section~\ref{RAD} presents our results from the analytic estimates of the dipole strength distribution for the 
three different $p$-wave one-neutron halos $^{31}$Ne and $^{34}$Na and $^{37}$Mg and also comparison is shown with FRDWBA results.
In Sec.~\ref{RAD2} the scaling of $S_n$ with dipole strength is discussed.
Finally, conclusions are made in Sec.~\ref{CON}.

\section{Electric-dipole strength distribution - The analytic set up}\label{MF}
We use an analytical model to calculate the
dipole strength distribution \citep{Nagarajan2005} for one neutron halo nuclei. 
The analytical model is first order theory and this theory works fairly well to study the nature of peripheral reactions involving the halo nuclei.
The dipole strength distribution for the single-particle dipole transition from a bound state $\phi_b(r)$ with $S_n$ to a 
continuum state $\phi_c(E_{c},r)$ with continuum energy $E_{c}$ is given by \citep{Bertulani1992, Nagarajan2005},
\begin{eqnarray}
\dfrac{dB(E1)}{dE_{c}} = (3/4\pi)(Z_{\rm eff}^{(1)}e)^{2} \langle\ell010|\ell^{'}0\rangle^{2}\nonumber\\ 
 \Bigg\vert\int \mathrm{d}r\ \phi_{b}(r)\ \phi_{c}(E_{c},r)r^{3}\Bigg\vert^{2}.
\end{eqnarray}
The effective charge $Z_{{\rm eff}}$ for a given multipolarity
${\lambda}$ is defined as \citep{Typel2005},
\begin{eqnarray}
Z_{\rm eff}^{(\lambda)} = {\frac{A_1^{\lambda}Z_2+(-1)^{\lambda}A_2^{\lambda}Z_1}{(A_1+A_2)^{\lambda}}},
\end{eqnarray}
with $Z_{i}$ and $A_{i}$ being the charges and masses of the core and of a neutron for $i= 1, 2$.
For the neutron dipole ($\lambda=1$) transitions $Z_{\rm eff}$ is simply reduced to $-Z/A$ and in the present study the spin is neglected for simplicity. 
Generally in two-body (core$+$neutron) model, the single-particle bound and continuum states are obtained as eigen states of the core$+$neutron potential.

In the limit of very low separation energy, the dipole strength distribution can be described analytically \citep{Bertulanii1988, Otsuka1994, Kalassa1996}. 
In fact, for the small $S_n$, most of the contribution in Eq.$~(1)$
comes from the asymptotic region, and therefore the asymptotic form of the wave function can be used for both the bound and 
continuum states. 
In the present study, the asymptotic form of the bound and continuum states are assumed to be expressed in terms of the spherical Hankel function of the first kind, $h_{\ell}^{(1)}(z)$,
with proper normalization and spherical Bessel function normalized to delta function, respectively.
The detailed mathematical expressions for these asymptotic forms of the wave functions  can be found in the Ref.~\citep{Nagarajan2005}.

By using these assumed asymptotic forms of the wave functions, Eq.$~(1)$ is reduced to much simpler form in the case of transitions starting from initial $p$-state. 
The simplified expression for the dipole strength distribution for the transition from the bound $p$-state to continuum $s$-state is given by,
\begin{eqnarray}
\dfrac{dB(E1)}{dE_{c}}(p \longrightarrow s)=\dfrac{\mu}{2\pi^2\hslash^2}(\hslash^2/2\mu)^{7/2}(Z_{\rm eff}^{(1)}e)^{2}N_b^2\nonumber\\
\dfrac{E_{c}^{1/2}(E_{c}+3S_{n})^{2}}{S_n^2(E_{c}+S_{n})^{4}},
\label{ps}
\end{eqnarray}
where $\mu$ is the reduced mass and $N_b$ is the normalization factor for the bound state.
For a dipole transition from bound $p$-state, the maximum of the strength distribution occurs at 
$E_{c}=0.18S_{n}$ for the continuum $s$-states.
Similarly for the transition from the bound $p$-state to continuum $d$-state the dipole strength distribution is given by,
\begin{eqnarray}
\dfrac{dB(E1)}{dE_{c}}(\textit{p}\longrightarrow \textit{d})=\dfrac{4 \mu }{\hslash^{2}\pi^{2}}(\hslash^{2}/2 \mu)^{7/2} (Z_{\rm eff}^{(1)}e)^{2}N_{b}^{2}\nonumber\\
 \dfrac{E_{c}^{5/2}}{S_{n}^{2}(E_{c}+S_{n})^{4}},
 \label{pd}
\end{eqnarray}
and for this case the maximum occurs at $E_{c}=5/3S_{n}$ for the continuum $d$-states.
The total B($E1$) values are obtained by integrating Eqs. (\ref{ps}) and (\ref{pd}) over the continuum energy ($E_c$). \\
One can also estimate the total B$(E1)$ in the extreme single-particle model provided other excited bound states do not contribute to the dipole transition, as
\begin{eqnarray}
B(E1)=\dfrac{3}{4\pi}(Z_{\rm eff}^{(1)}e)^{2}\langle r^2 \rangle.
\label{be}
\end{eqnarray} 
Expressing the bound state wave function ($\phi_b$) in terms of the asymptotic form of the spherical Hankel function,  
 the mean square radius of bound $p$-state is simply given by,
\begin{eqnarray}
\langle r^2 \rangle = \langle \phi_b(r) \vert r^2 \vert \phi_b(r) \rangle.
\end{eqnarray}
Taking, 
\begin{eqnarray}
\phi_b(r)=N_b h_{1}^{(1)} (i a r)  \xrightarrow{r \to \infty}
 N_{b}\ i\ exp(-ar)/ar,
\end{eqnarray}


where, $a^2=\dfrac{2\mu S_n}{\hbar^2}$, the mean square radius further reduces to the simpler form 
\begin{eqnarray}
\langle r^2 \rangle_{MI}= \dfrac{1}{2a^2}
= \dfrac{\hslash^2}{4\mu S_n}.
\label{sn}
\end{eqnarray}
This method of obtaining total B($E1$) value can be called as model independent method (MI). From Eqs.~(\ref{be}) and (\ref{sn}) 
it can be found that the total B($E1$) strength is inversely proportional to the separation energy \citep{Nakamura2012} in actual cases. Interestingly taking the asymptotic form of the Hankel function itself makes $\langle r^2 \rangle_{MI}$ independent of $\ell$.

The Coulomb breakup reactions with halo nuclei provides a convenient tool to study their electric-dipole responses \citep{Nakamura1994, Nakamura1999, Baur2003}.
In order to compare the analytic estimation of dipole strength distribution with the realistic reaction theory estimate 
we have used  the post form finite-range distorted-wave born approximation (FRDWBA) theory \citep{Chatterjee2000}.

In FRDWBA framework, for the Coulomb breakup reaction, $a + t$ $\rightarrow$ $d + n + t$, 
where the projectile (a neutron halo) breaks up into two fragments $d$ (charged core) and $n$ (valence neutron) in the Coulomb field of a target $t$, 
the triple differential cross-section is given by
\begin{eqnarray}
\dfrac{d^3\sigma}{dE_{n} d\Omega_b d\Omega_n} = \dfrac{2 \pi}{\hbar \upsilon_{at}} \rho \sum_{\ell m} \vert \beta_{\ell m} \vert^2 .
\label{DWBA1}
\end{eqnarray} 
Here $\upsilon_{at}$ is the relative velocity of $a-t$ system in the initial channel, 
$\rho$ is the appropriate three-body phase space factor \citep{Fuchs1982} and $\beta_{\ell m}$ is the reduced transition matrix. Eq.~(\ref{DWBA1}) is multiplied with appropriate jacobian to get the relative energy spectra $\dfrac{d\sigma}{dE_{c}}$ \citep{Fuchs1982, Banerjee2008}. 
Input to the FRDWBA theory is the realistic wave function that describes the relative motion between the
valence neutron and the core in the ground state of the projectile.
The detailed formalism and discussion of FRDWBA can be found in \citep{Chatterjee2000, Chatterjee2018}.
The Coulomb breakup cross-section is related to the dipole strength distribution [$dB(E1)/dE_{c}$] and is given by relation \citep{Bertulani1988,Nakamura1999},
\begin{eqnarray}
\dfrac{d\sigma}{dE_{c}} = \dfrac{16 \pi^3}{9 \hbar c} n_{E1}\dfrac{dB(E1)}{dE_{c}},
\label{DWBA2}
\end{eqnarray}
where $n_{E1}$ is the number of virtual photons for $E1$ excitation. 
We will compare our analytic estimate for the dipole strength distribution with the one obtained by Eq.~(\ref{DWBA2}). It needs to be mentioned, of course, that the calculation of dipole strength [$dB(E1)/dE_{c}$] would be independent of the beam energy of the projectile. This gives us the opportunity to relate our relative energy spectra to those already reported in Refs. \citep{Shubhchintak2014,Singh2016,Shubhchintak2015}.

As discussed in \citep{Nagarajan2005}, one expects the deviation in the analytic estimate from the exact calculations, i.e., FRDWBA estimate. 
This deviation originates from the effect of the binding potential on the bound wave function. 
The multiplicative correction factor (CF) can be obtained for this deviation.
In order to derive the CF we assume a realistic potential, i.e., Woods-Saxon potential (WS) with radius parameter as $1.236$\, fm and diffuseness parameter as $0.62$\,fm. 
The CF depends on the separation energy and the potential radius $R$, and is given by
\begin{eqnarray}
{\rm CF} = \dfrac{F(S_n, R)}{\langle r^2 \rangle_{MI}},
\end{eqnarray}
with $F(S_n,R)$ as 
\begin{equation}
F(S_n,R) = \dfrac{A_{d}}{A_{a}} \langle r^2 \rangle_d+\dfrac{A_{d}*A_{n}}{A_{a}^2} \langle r^2 \rangle_{\rm WS}. 
\label{cf}
\end{equation}
In Eq.~(\ref{cf}), $A_{d}$ is the mass of the core, $A_{n}$ is the mass of the valence neutron and $A_{a}$ is the mass of the projectile. The mean square radius of the core is $\langle r^2 \rangle_d$ and $\langle r^2 \rangle_{WS}$ is the mean square intercluster distance between core and valence neutron using the Woods-Saxon potential. 
Eq.~(\ref{cf}) is a consequence of considering the finite size of the core of the nucleus \footnote{The mean square matter radius of a nucleus with mass number A in terms of expectation values, (see Eq. (A.4) in Ref.\citep{Mason2009})
\begin{equation}
\langle r^2 \rangle_{A_1+A_2}=\dfrac{A_1}{A}\langle r^2 \rangle_{A_1}+\dfrac{A_2}{A}\langle r^2 \rangle_{A_2}+\dfrac{A_1 A_2}{A^2}\langle R^2 \rangle \nonumber
\label{cf1}
\end{equation}
In our case $A_1=A_d$, $A_2=A_n$, $A=A_a$ and we have neglected the second term of above contributed by the neutron.}.

 
In the next section, the analytic estimates of Eq.~(\ref{ps}) and Eq.~(\ref{pd}) with CF has been compared with FRDWBA estimates.

\section{Results and Discussions}\label{RAD}
We study the dipole strength distribution for the three $p$-wave one neutron halos, i.e., $^{31}$Ne, $^{34}$Na and $^{37}$Mg at zero deformation. 
The total dipole strength for all these three systems include contributions from two transitions, i.e., from the ground $p$-state to continuum $s$ and $d$-states. The parameters used in Eq.~(\ref{cf}) for the three nuclei are summarized in Table 1.

\begin{table}[ht]
\caption{The parameters used in Eq.~(\ref{cf}) are $r_d$ = $\sqrt{\langle r^2 \rangle_d}$ = $1.236 A_{d}^{1/3}$ and $r_{WS}$ = $\sqrt{\langle r^2 \rangle_{WS}}$ (calculated from WS potential), for the three nuclei.}
\centering
\begin{tabular}{ |c|c|c|c| } 
\hline
Nucleus & $r_d$ (fm) &  $r_{WS}$ (fm)\\
\hline
~~~~~~~$^{31}$Ne~~~~~~~ & ~~~~~~~3.84~~~~~~~ & ~~~~~~~3.98~~~~~~~ \\ 
~~~~~~~$^{34}$Na~~~~~~~ & ~~~~~~~3.96~~~~~~~ & ~~~~~~~4.14~~~~~~~ \\ 
~~~~~~~$^{37}$Mg~~~~~~~ & ~~~~~~~4.09~~~~~~~ & ~~~~~~~4.19~~~~~~~\\ 
\hline
\end{tabular}

\end{table}

\begin{figure}[h!]
\centering
\includegraphics[height=8cm,width=9cm]{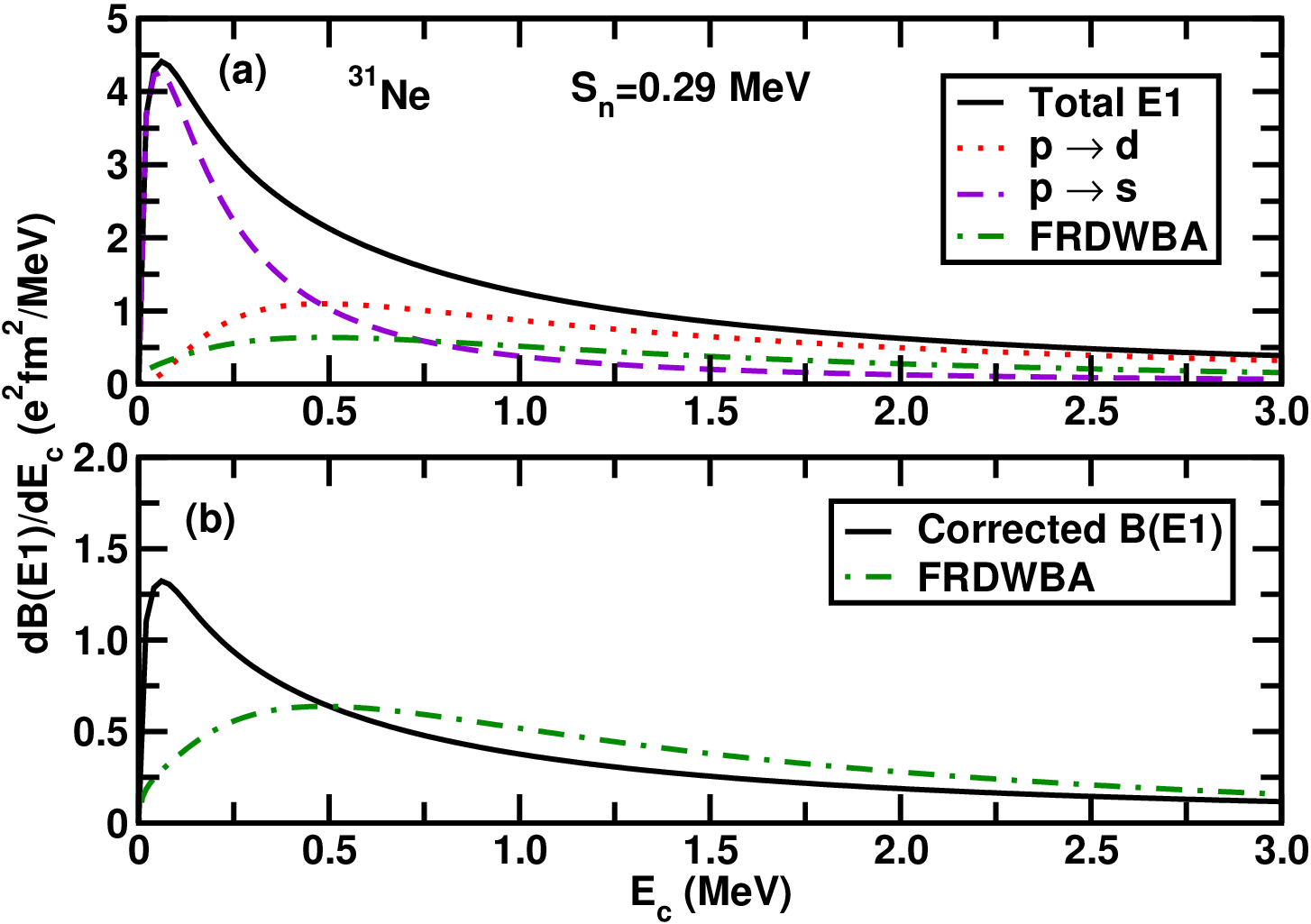}
\caption{(Color online) Dipole strength distribution in $^{31}$Ne as a function of the energy in the continuum, for the initial $p$- state bound by $S_n=0.29$\,MeV. 
(a) Solid curve refers to the analytical results without CF for the total dipole strength, where as the dashed and dotted curve refers to the transitions from 
the ground $p$-state to the continuum $s$ and $d$-states respectively. The dot-dashed curve refers to the FRDWBA results. (b) Comparison of the analytic estimate with the 
CF and the FRDWBA results.}
\label{Ne}
\end{figure}

Figure~\ref{Ne} plots the dipole strength distribution for $^{31}$Ne with the continuum energy ($E_c$). 
The percentage contribution for the transition from the ground $p$-state to continuum $s$ and $d$-states are $48\%$ and $52\%$ respectively.
The maximum of the strength occurs at $E_{c}=0.05$\,MeV which is $0.18$\,$S_n$ for $p$ to $s$-state and at $E_{c}=0.48$\,MeV which is $5/3$\,$S_n$ for $p$ to $d$-state.
Figure~\ref{Ne}(a) shows the the analytic estimate without CF for the total dipole strength and its contributed transition strengths.
In order to compare the analytic estimate of dipole strength distribution we study the Coulomb dissociation of $^{31}$Ne on $^{208}$Pb at $234$\,MeV/u. 
Figure~\ref{Ne}(a) also shows the comparison of FRDWBA estimate with the total dipole strength obtained analytically without CF, 
we find the noticeable difference between both of these estimates. 
As discussed in section~\ref{MF}, one needs to introduce the multiplicative CF for the analytic estimates.
Figure~\ref{Ne}(b) shows the comparison between the analytic results with CF$=0.3$ and the FRDWBA results. 
The grazing angle for FRDWBA calculation is $1.1\degree$ which ensures the 
dominance of Coulomb breakup. 
The total integrated dipole strength for $^{31}$Ne case amounts to be $1.3$ ${e}^2\rm{fm}^2$ and $1.1$ ${e}^2\rm{fm}^2$ from the analytic and FRDWBA estimates respectively, whereas 
from the model independent calculation it amounts to be $1.3$ ${e}^2\rm{fm}^2$.
\begin{figure}[h!] 
\centering
\includegraphics[height=8cm,width=9cm]{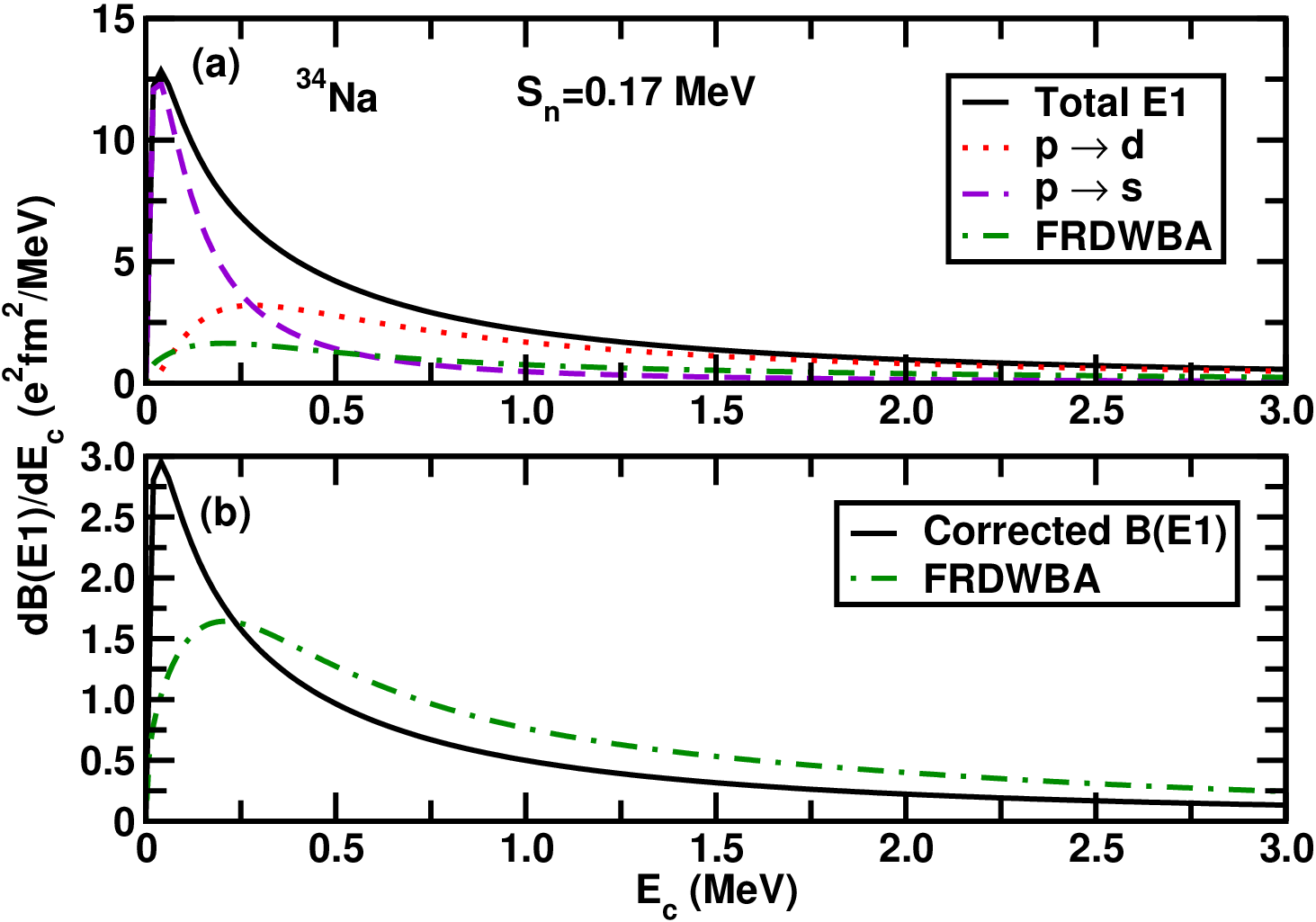}
\caption{(Color online) Dipole strength distribution in $^{34}$Na as a function of the continuum energy, for the initial $p$- state bound by $S_n=0.17$\,MeV. 
(a) Solid curve refers to the analytical results without CF for the total dipole strength, where as the dashed and dotted curve refers to the transitions from 
the ground $p$-state to the continuum $s$ and $d$-states respectively. The dot-dashed curve refers to the FRDWBA results. (b) Comparison of the analytic estimate with the 
CF and the FRDWBA results.}
\label{Na}
\end{figure}

Figure~\ref{Na}(a) shows the total dipole strength distribution of $^{34}$Na obtained analytically without CF along with its comparison with FRDWBA estimate. Similar to 
$^{31}$Ne we find the noticeable difference between both of these estimates.
The transitions from the ground $p$-state to the continuum $s$ and $d$-states contribute $44\%$ and $56\%$ respectively to the total strength. 
We observe a maximum of the strength distribution occurs at $E_{c}=0.03$ MeV which is $0.18$\,$S_n$ for $p$ to $s$-state and at $E_{c}=0.28$ MeV which is $5/3$\,$S_n$ for $p$ to $d$-state.
We study the Coulomb dissociation of 
$^{34}$Na on $^{208}$Pb at $100$\,MeV/u to compare the analytic estimate of dipole strength distribution.
The comparison between the analytic results with CF$=0.23$ and the FRDWBA results shown in Fig.~\ref{Na}(b) and the grazing angle for FRDWBA calculation is $2.1\degree$. 
From both analytic and FRDWBA theory, the total integrated dipole strength for $^{34}$Na case comes to be $1.7$ ${e}^2\rm{fm}^2$ and $2.0$ ${e}^2\rm{fm}^2$ respectively, whereas 
from the model independent calculation it comes to be $2.2$ ${e}^2\rm{fm}^2$.
\begin{figure}[h!]
\centering
\includegraphics[height=8cm,width=9cm]{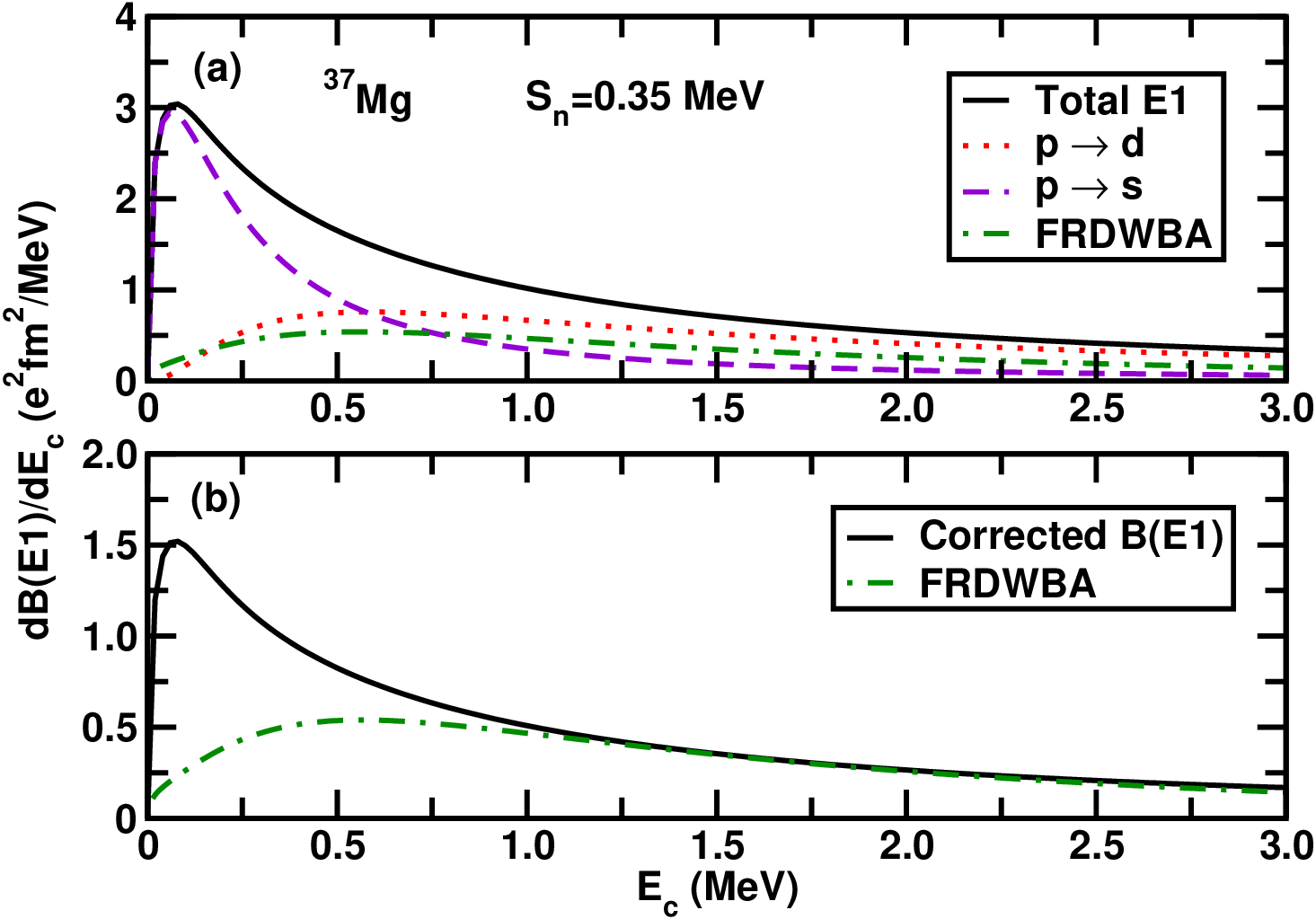}
\caption{(Color online) Dipole strength distribution in $^{37}$Mg as a function of the continuum energy, for the initial $p$- state bound by $S_n=0.35$\,MeV. 
(a) Solid curve refers to the analytical results without CF for the total dipole strength, where as the dashed and dotted curve refers to the transitions from 
the ground $p$-state to the continuum $s$ and $d$-states respectively. The dot-dashed curve refers to the FRDWBA results. (b) Comparison of the analytic estimate with the 
CF and the FRDWBA results.}
\label{Mg}
\end{figure}

Similar to $^{31}$Ne and $^{34}$Na Fig.~\ref{Mg}(a) shows the dipole strength distribution of $^{37}$Mg obtained analytically without CF 
along with its comparison with FRDWBA estimate, again we find the noticeable difference between both of these estimates.
The transitions from the ground $p$-state to the continuum $s$ and $d$-states contribute $51\%$ and $49\%$ respectively to the total strength.  
For $p$ to $s$-state we obtain a maximum of the strength distribution at $E_{c}=0.06$\,MeV which is $0.18$\,$S_n$ and for $p$ to $d$-state, it is at $E_{c}=0.58$\,MeV 
which is $5/3$\,$S_n$.
In order to compare the analytic estimate of dipole strength distribution we study the Coulomb dissociation of 
$^{37}$Mg on $^{208}$Pb lead target at $244$\,MeV/u.
Figure~\ref{Mg}(b) shows the comparison between the analytic results with CF$=0.5$ and the FRDWBA results with grazing angle equals to $1.1\degree$. 
For $^{37}$Mg, the total integrated dipole strength amounts to be $1.4$ ${e}^2\rm{fm}^2$ and $1.0$ ${e}^2\rm{fm}^2$ from both the analytic and FRDWBA estimates respectively, whereas 
from the model independent calculation it amounts to be $1.1$ ${e}^2\rm{fm}^2$.

The analytic model which uses the asymptotic form of the bound and continuum wave function only gives a broad trend of the observable. 
The FRDWBA model on the other hand includes more realistic wave function. 
Therefore the disagreement at low energies which calculates $\dfrac{dB(E1)}{dE_c}$ should be seen in this light. 
We however reiterate that the total B($E1$) is similar in magnitude for the analytic, FRDWBA and the so called model independent estimates. 
However it is not clear if this similarity is due to a particular separation energy or it should be valid even if 
the separation energy is varied. In fact, this gives us the motivation to probe the variation of the 
low lying dipole strengths in these nuclei as a function of its separation energy.\\

\subsection{Scaling of  the dipole strength with separation energy}\label{RAD2}

\begin{figure}[h!]
\centering
\includegraphics[height=8cm,width=9cm]{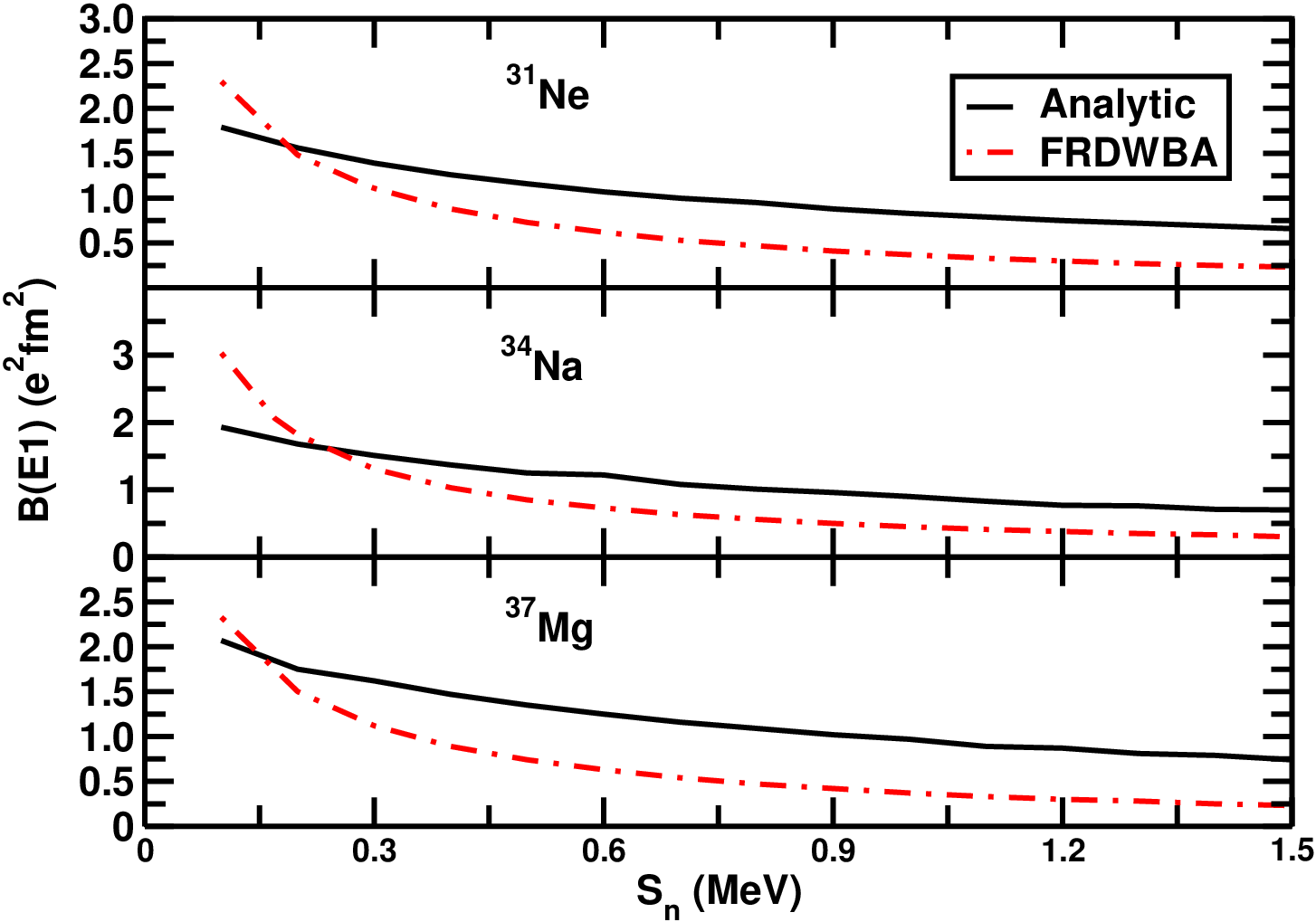}
\caption{Total B(E1) strength of $^{31}$Ne, $^{34}$Na and $^{37}$Mg as a function of separation energy. The dot-dashed line shows the FRDWBA contribution where we adjust the Woods-Saxon parameters 
to obtain the correct separation energy of the projectile and solid line is the analytical calculation.}
\label{M}
\end{figure}

We make analytic and FRDWBA estimates of the total integrated dipole strengths (B($E1$)) for $^{31}$Ne, $^{34}$Na and $^{37}$Mg 
with wide range of the separation energies, i.e., $0.1$-$1.5$\,MeV. 
Figure~\ref{M}, plots the B($E1$) as a function of the separation energy. 
The dot-dashed line in the Fig.~\ref{M} refers to the FRDWBA results, where bound and continuum single-particle states 
are generated by using a Woods-Saxon potential and the potential parameters are adjusted to reproduce the separation 
energy of the ground state of projectile. The solid line refers to the analytic estimate. 
It is interesting to estimate the separation energy at the point of intersection of the analytic and FRDWBA results, and compare it with available data.

For $^{31}$Ne, our estimated separation energy value is $0.18$\,MeV which is consistent with the measured value ($0.15^{+0.16}_{-0.10}$\,MeV) 
via $1n$-removal reactions \citep{Nakamura2014}. On other-hand for $^{34}$Na our estimated separation energy is $0.24$\,MeV and the 
measured value ($0.17\pm50$\,MeV) \citep{Gaudefroy2012} has large error bars but still our estimated value is consistent within these error bars 
along with other theoretical estimates \citep{Fortune2013}. For $^{37}$Mg our estimated separation energy is $0.15$\,MeV which is also consistent with the 
measured value ($0.22^{+0.12}_{-0.09}$)\,MeV \citep{Kobayashi2014}. This method could thus be useful for making an initial estimates of the 
separation energy of weakly bound exotic nuclei, where measured data are rare.

\section{Conclusion}\label{CON}

In summary, the dipole strength distribution has been studied for three probable $p$-wave one-neutron halos $^{31}$Ne, $^{34}$Na and $^{37}$Mg 
lying in the island of inversion by using two different approaches, i.e., with an analytic model and with the FRDWBA theory. 

One of the problems in a first order theory is to fix the position and widths of continuum states. 
The first order analytical model neatly overcomes this, at least for peripheral reactions, by replacing the continuum wave 
function with that of an universal asymptotic type for the case of a loosely bound neutron halo this is approximately a spherical bessel function. On the other hand the FRDWBA, derived from a post-form reaction theory framework, have contributions from the entire continuum 
corresponding to all the multipoles and the relative orbital angular momenta. The ground state wave function of the projectile is the 
only input to FRDWBA. The relative energy spectra in the Coulomb dissociation of $^{31}$Ne, $^{34}$Na and $^{37}$Mg on $^{208}$Pb at $234$, $100$ and $244$\,MeV/u, 
respectively was used as a base to calculate the dipole strength distribution, in the FRDWBA theory.

With a correction factor to the analytical model, derived from an realistic potential, comparison with the FRDWBA results 
showed reasonably good agreement, at least at the total dipole strength level. This led us to study the scaling of the total B(E1) strengths with the one neutron separation energy.
Apart from analyzing the peak of the dipole strength distribution and the relative energy spectra we believe our method of 
comparing the total total dipole strengths could also lead to a good estimate of the one-neutron separation energies for the $p$-wave halos. 
Our estimated separation energy values are $0.18$, $0.24$ and $0.15$\,MeV for $^{31}$Ne, $^{34}$Na and $^{37}$Mg, respectively. 
These values are in good agreement with the experimental data, wherever available, and the previous theoretical estimates.

Finally, let us also mention that nuclear breakup effects cannot be completely eliminated although dominance of Coulomb breakup can be ensured in experiments by considering forward angle scattering. A proper estimation of nuclear breakup contributions would require knowledge of optical potentials which are difficult to obtain in this region. For the case of a light exotic nucleus like $^{11}$Be breaking up on heavy target ($^{197}$Au), the nuclear breakup contribution to the total one neutron removal cross section was estimated \citep{Chatterjee2003} to be of the order of 10$\%$. So, grossly we believe that for the breakup of exotic nuclei in the medium mass region on a heavy target the nuclear contribution would be of a similar magnitude. This would also be an interesting case to investigate further. 

The present study, also, ignores the deformation effects in the ground state of these $p$-wave halo nuclei for the time being. 
The calculations with two dimensional scaling, including deformation, are in the progress and will be reported in the near future.


\section*{Acknowledgments}
Manju acknowledges financial support from MHRD (Govt. of India) for a doctoral fellowship.
J. Singh gratefully acknowledged the financial support from Nuclear Reaction Data Centre (JCPRG), Faculty of science, Hokkaido University, Sapporo, Japan.

\end{document}